# Next generation portal for federated testbeds

MySlice v2: from prototype to production

Loïc Baron, Radomir Klacza, Pauline Gaudet-Chardonnet, Amira Bradai, Ciro Scognamiglio, Serge Fdida
Network Performance Analysis department (NPA),
LIP6 (UMR 7606), Sorbonne Universities (UPMC),
Paris, France

*Abstract*—A number of projects in computer science around the world have contributed to build federated experimental facilities providing access to a large set of distributed compute, storage and network resources for the research community. Several tools have been developed to provide users an easy access to the federated testbeds. This paper presents the architecture of the new version of the MySlice web portal, that has evolved from a prototype to a production ready software.

I. INTRODUCTION

The technical challenge behind a portal for federated testbeds is that resources are heterogeneous and under the control of multiple authorities. In addition, to build a federated system, experimenters at one institution must be authenticated and authorized to gain access to testbed resources hosted by other authorities. The Slice-based Federation Architecture (SFA) [1] addresses this authentication and authorization challenge.

Above these APIs, several client tools [2] have been developed including command line (omni, sfi) and graphical tools (Flask, jFed [3], MySlice). But the large number of distributed platforms tend to increase both the number of possible points of failures and the delay caused by the long response time of some platforms.

The first prototype of MySlice v1 [4] was tightly coupled with the underlying AM API [5]. As long as one of the Aggregate Managers of the federation was down or answered with a long delay, the user experience was affected. Moreover, a large number of redundant queries were sent at the same time by different users, such as the list of resources. These results can be stored in a database in order to avoid multiple calls to the AMs. The version 2 of MySlice addresses those issues improving thus the user experience.

As a production ready software, the objectives of MySlice V2 are threefold:

- **Reliability:** the architecture decouples the APIs and the Web. Each component is developed and tested separately to ensure the maintainability of relatively small components. Unit tests have been designed to be executed after each commit of the code using GitLab Continuous Integration ensuring that the released version of the code is compliant. This version also improves errors and exceptions handling and their reporting in log files. Finally, a monitoring tool has been developed to run the full experiment lifecycle through the MySlice v2 REST API every 15 minutes.

- **Scalability:** The core architecture has been tested in the context of a MOOC with thousands of users. Therefore, we are confident that it can handle a heavy simultaneous usage. Each long-lasting query sent to the system is an event placed in a queue. It allows to asynchronously process the events that trigger the AMs API using MySlice Lib. The queues can have multiple threads allowing the system to scale up.

- **Usability:** The Web User Interface has been redesigned to be faster and more user-friendly. The web UI interacts with a document database through a REST API and a Websocket ensuring thus fast and interactive results. The User Interface has been enriched with documentation about the testbeds and nodes mapping.

II. MYSLICE V2

A. Architecture

The new architecture is composed of 5 layers with a clear separation of concerns: Web frontend, APIs (REST/WS), Database, Services with workers and Library (XML-RPC), see Annex 1.

The frontend has been redesigned using the ReactJS framework. The benefit of using such a framework is to create generic components that can be re-use in different views depending on the properties passed to the components. Moreover, the management of a store that maintains a state of a component or a view is very well suited for an event-oriented application.

We have clearly defined the REST and WebSocket APIs used by the React components and third-party software. The web components are able to get or post data through the REST API and can be notified of a change through the WebSocket, providing a very interactive frontend.

Some interactions of the user with the frontend are generating events that are stored in a document oriented database. The MySlice router is then responsible to place these events in the relevant queue depending on their type. Each type of event is asynchronously processed by a service. The services are calling workers that can be multithreaded to scale up the capabilities of the system. The workers are responsible of the interactions with the distributed testbeds through the AM API (XML-RPC) and with the SFA Registry, which is the root authority of the federation providing the credentials to access the testbeds.

*B. Performances*

MySlice portal offers an interface to the root of trust of the federation (SFA Registry) allowing users to register, to create a project in which he/she can collaborate with other experimenters and to create an experiment (called a Slice) in order to reserve resources. The communication between MySlice v2 and the SFA Registry has been improved. For instance, the creation of a project takes only an average of 1.3 seconds and the creation of a slice an average below 2 seconds, as shown in Figure 1 and Figure 2.

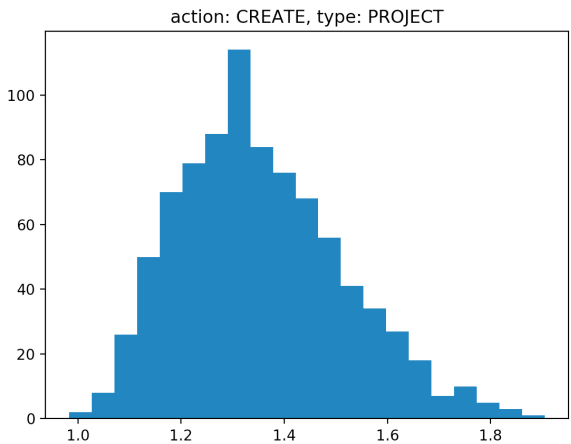

**Figure 1: Create project events**

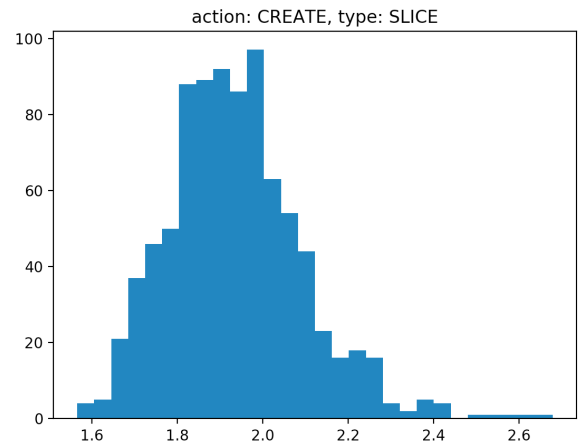

**Figure 2: Create slice events**

As presented in the architecture, MySlice v2 features a database that stores the information about the objects of the federation as JSON documents. Synchronization processes periodically refresh the data of this caching system. Depending on the object, the periodicity ranges from once a day for the list of authorities at the SFA Registry to once every 5 minutes for the list of leases at the AMs that support scheduled reservations of resources.

As an example, to illustrate the improvement of the performances for the end-user, the call to list the resources using the AM API can last for 10 seconds to a minute. Whereas the call to the MySlice v2 database takes only an average of 0.20 seconds as shown in Figure 3.

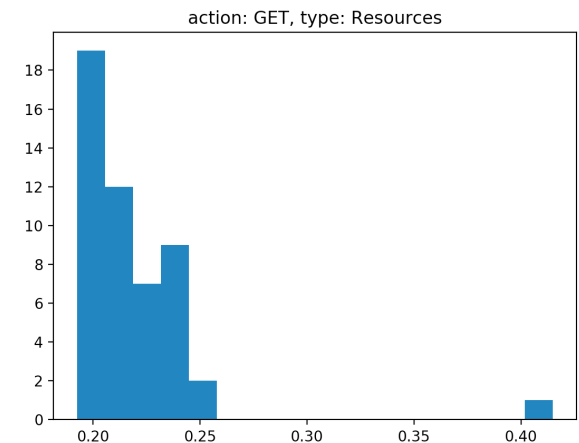

**Figure 3: Get the list of resources using REST API**

Of course, the improvement of the performances is limited to the boundaries of the MySlice v2 internal architecture and can't mitigate some delays inherent to the remote testbeds, for instance during the reservation process. As a result, the scheduled reservation of resources takes an average of 45 seconds as shown in Figure 4.

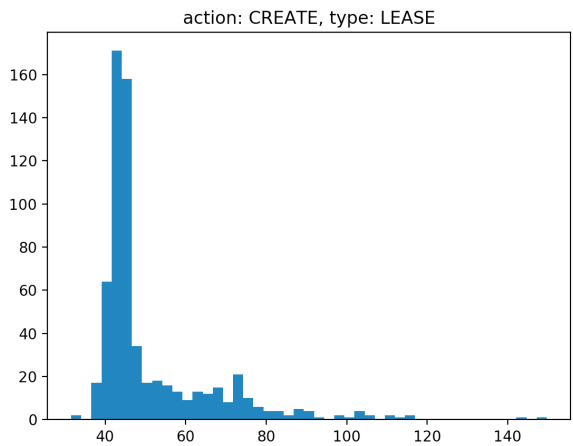

**Figure 4: Create lease events, scheduled reservation of resources**

## III. USE-CASE AND DEMO

The OneLab federation [6] is composed of the following testbeds:

- **NITOS** operated by the University of Thesaly (UTH) provides wireless nodes based on Wi-Fi, WiMax and LTE technologies. It offers 3 environements: outdoor, indoor RF isolated and indoor office [7].
- **FIT wireless Paris** hosted by Université Pierre et Marie Curie (Sorbone Universités) provides 40 WiFi nodes in an indoor environment.
- **FIT R2lab** hosted by INRIA in Sophia Antipolis offers 37 WiFi nodes in an anechoic chamber along with USRP and LTE nodes [8].
- **FIT IoT-Lab** offers six sites scattered around France with a total of 2728 wireless sensor nodes for testing Internet of Things [9].
- **PlanetLab Europe** provides more than 300 virtual container nodes on many sites worldwide, allowing for large-scale, extensive, and remote testing [10].

The demo will walk the attendees through the process of creating an account, creating a project, reserving resources. Users will be able to sign up and their accounts will be authorized on the spot, so that he/she may immediately access the portal resources and deploy experiments.

## IV. CONCLUSION AND FUTURE WORK

The reliability, the scalability and the usability of MySlice v2 have been improved through a brand-new architecture, that features asynchronous events processing and a database caching system. As shown in this paper, the performances have been improved. The software is able to deliver very quickly a relatively up-to-date list of the available federated resources. It provides thus a smooth user experience. While trying to mitigate the delay, the reservation process remains under the responsibility of the distributed testbeds. A future work should focus on the redesign of the AM API using modern technologies, such as REST or AMQP instead of XML-RPC. This would greatly improve the overall performance of the federated system.

MySlice v2 is composed of different plugins that can be independently developed by different contributors. This modular architecture allows its extension on both sides either to support new backend services or to enhance the web frontend. The modularity of the architecture ensures the sustainability of the solution, as new modules can be developed to cope with the evolutions of the federated platforms.

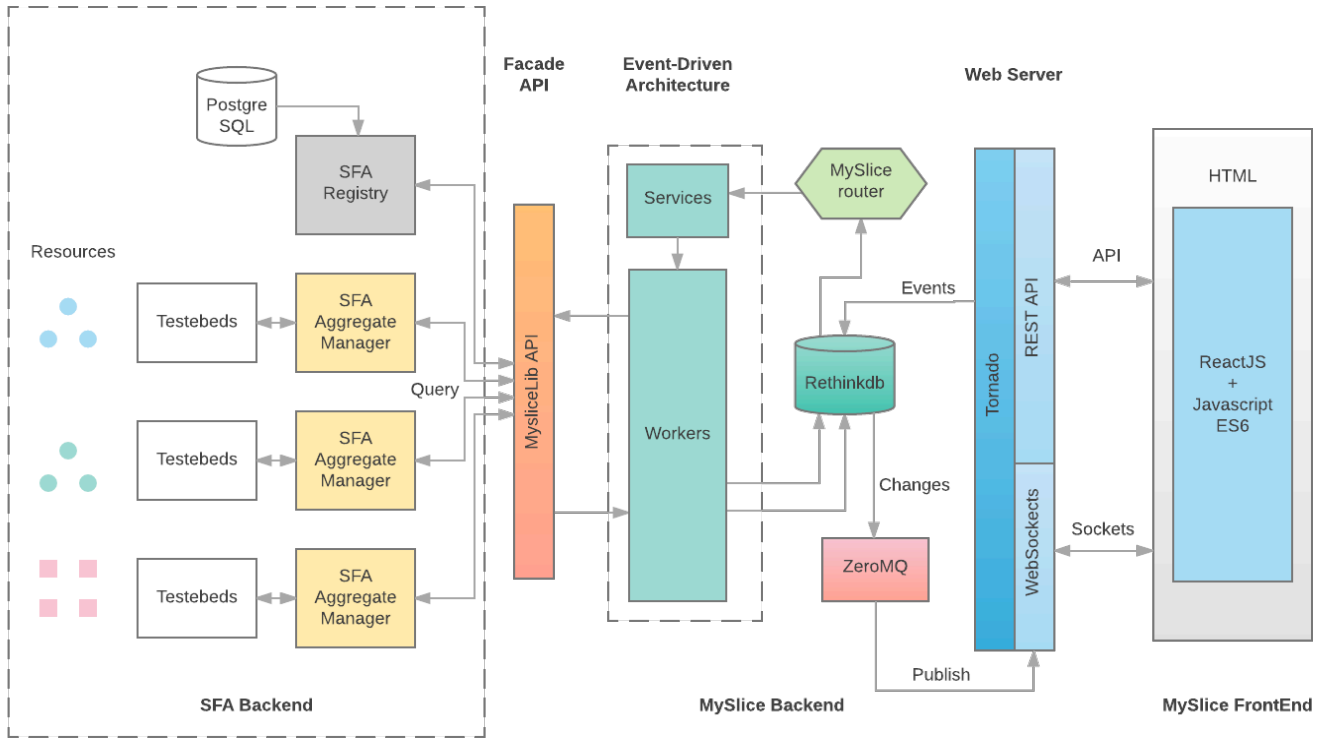

**Annex 1: MySlice v2 architecture**